\documentclass[acmtog, nonacm]{acmart}

\usepackage{booktabs} %

\usepackage[ruled]{algorithm2e} %

\SetAlFnt{\small}
\SetAlCapFnt{\small}
\SetAlCapNameFnt{\small}
\SetAlCapHSkip{0pt}

\usepackage{xspace}

\makeatletter
\DeclareRobustCommand\onedot{\futurelet\@let@token\@onedot}
\def\@onedot{\ifx\@let@token.\else.\null\fi\xspace}

\def\eg{\emph{e.g}\onedot}

\makeatother

\acmJournal{TOG}

\newcommand{\bs}[1]{\boldsymbol{#1}}

\begin{document}
\title{From Universal Humanoid Control to Automatic Physically Valid Character Creation}

\author{Zhengyi Luo}
\author{Ye Yuan}
\author{Kris M. Kitani}
\affiliation{%
 \institution{Carnegie Mellon University}
  \country{USA}
}

\begin{abstract}
Automatically designing virtual humans and humanoids holds great potential in aiding the character creation process in games, movies, and robots. In some cases, a character creator may wish to design a humanoid body customized for certain motions such as karate kicks and parkour jumps. In this work, we propose a humanoid design framework to automatically generate physically valid humanoid bodies conditioned on sequence(s) of pre-specified human motions. First, we learn a generalized humanoid controller trained on a large-scale human motion dataset that features diverse human motion and body shapes. Second, we use a design-and-control framework to optimize a humanoid's physical attributes to find body designs that can better imitate the pre-specified human motion sequence(s). Leveraging the pre-trained humanoid controller and physics simulation as guidance, our method is able to discover new humanoid designs that are customized to perform pre-specified human motions. Video demos can be found at \href{https://zhengyiluo.github.io/projects/agent_design/}{the project page }.
\end{abstract}



\begin{teaserfigure}
  \includegraphics[width=\textwidth]{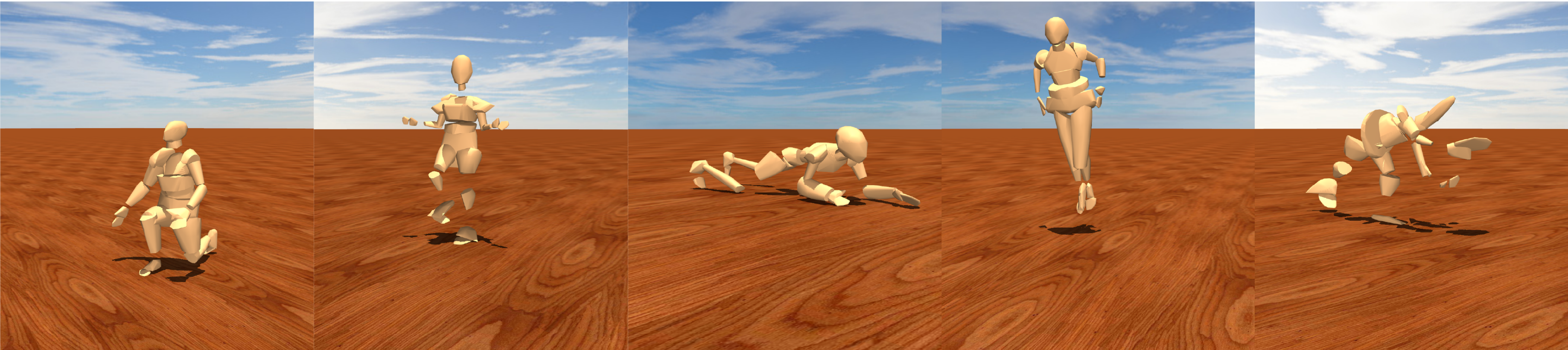}
  \caption{Discovered designs for parkour, jump-rope, crawl, jumping jacks, and cartwheel.}
  \label{fig:teaser}
\end{teaserfigure}

\maketitle

\section{Introduction}

We consider the task of determining the body structure of a digital character that possesses the physical qualities necessary to perform a certain range of tasks. For example, a digital character that needs to perform cartwheels or back-flips might require a light body, or a character that needs great balance might have a heavier lower body and a lower center of mass. To generate such digital characters whose physique is representative of the type of actions it must perform, we proposed a method for automatically generating novel humanoid designs that are physically valid and conditioned to perform a set of predefined actions.

Concretely, given the sequence(s) of human motion captured by motion capture, our method will output a humanoid design that can successfully imitate that sequence in a physics simulator (Fig. \ref{fig:teaser}). 
To find humanoid designs that are suited for performing a diverse set of motions, we utilize physics simulation and data-driven motion imitation techniques to enforce physical plausibility and human likeness. Grounded in physical laws, our method can recover character designs that can serve as blueprints to aid the ideation process for creating uniquely looking yet physically controllable characters. 
 
The challenges are two-fold: (1) the design space to search for suitable characters is immense: (2) we desire our recovered character design to specialize in performing the input motion while staying useful for general daily motion tasks (\emph{e.g.}, walking and running). However, controlling a humanoid to imitate diverse reference motions inside the physics simulation itself is a challenging and time-consuming task \cite{Mourot2021ASO}. Start-of-the-art methods often take days or weeks to train suitable controllers for a \textit{single} character design \cite{Won2020-lb, wang2020unicon, Fussell2021SuperTrack, yuan2020residual, Luo2021DynamicsRegulatedKP}. This can be attributed to the difficulty of controlling a simulated character with such high degrees-of-freedom (e.g., 69 on the SMPL \cite{Loper2015SMPLAS} human model) and the dynamics mismatch between simulated rigid bodies and real-world physics. As a result, while algorithmic discovery of agent design and morphology has been applied to simple walkers and hoppers \cite{Wang2019NeuralGE, Luck2019DataefficientCO, Yuan2021Transform2ActLA} (such as those defined in the popular OpenAI Gym \cite{Brockman2016OpenAIG}), learning specialized humanoid designs given a reference motion has not yet been attempted. 

In this paper, we tackle automatic and efficient character discovery from two complementary angles: (1) learning a general character controller that can perform tens of thousands of human motions, given reference motions and pre-scanned human physique; (2) given a sequence(s) of reference motion, we form a dual design-and-control policy for intelligently altering the character's design. Specifically, we first incorporate the human body shape into the state space of our controller and learn from a large-scale human motion dataset \cite{Mahmood2019AMASSAO} accompanied by realistic human body shape annotation. 
Based on the key observation that foot geometry and compliance are key contributors to agent instability \cite{Park2018MultiSegmentFM} (29 muscles associated with the soft and compliable human foot vs 2 pieces of simple rigid bodies in most of the simulated characters), we propose a simple yet effective way to compensate for such a mismatch. We argue that while the multi-joint foot proposed in \cite{Park2018MultiSegmentFM} is anatomically correct, humans wear shoes and footwear that are difficult to perfectly simulate. Instead of adding additional joints to model a flexible foot, we apply residual external forces \cite{yuan2020residual, Yuan2021SimPoESC, Luo2021DynamicsRegulatedKP} \textit{on the foot}. We also limit the forces to only be applied when the foot is in contact with the ground to maximize its physical realism. In this way, we account for the mismatch between the simulated humanoid and real human foot, while still being able to imitate tens of thousands of human motion sequences.

After the controller is trained, we compose a dual design-and-control framework for automatic character discovery based on reference motion. Here, we extend the recently proposed Transform2Act \cite{Yuan2021Transform2ActLA} framework to a multitask setting for the challenging humanoid motion imitation task. Although Transform2Act has been shown to be effective in finding a \textit{single} agent design suitable for \textit{one} specific tasks, it is difficult to directly apply it to the 3D human motion imitation process. Unlike tasks such as \textit{move forward}, motion imitation is much more multimodal ---- different types of human motion demand unique agent designs. In addition, simple walkers and hoppers can freely change their geometry, but the human body follows a predefined biological structure and movement pattern that require additional constraints. Based on these observations, we utilize a pre-learned controller as the control policy and the dynamics prior to guide the agent discovery process. Specifically, we pair the pre-trained controller with a design policy to vary the agent's design and receive feedback based on motion imitation quality. During this process, we keep the control policy \textit{fixed} to ensure that the newly found character can still be controlled by the fixed control policy. Intuitively, we are looking for characters that are within the controllable character range learned from a large database of realistic human motion and body shapes. 

In summary, the main contributions of this work are: (1) We propose an automatic physically valid humanoid character discovery framework that can find character design based on target human motion sequences; (2) We propose a generalized controller that can control characters of varying body shapes to imitate tens of thousands of human motions; (3) Experiments show that our approach can not only identify character designs that are unique and improve the motion imitation quality for the given motion sequences, but also remains capable of imitating a large corpus of human motion.

\section{Related Works}
\subsection{Simulated Character Control}

Controlling a simulated character inside a physics simulation \cite{Raibert1991AnimationOD, Peng2018DeepMimicED, wang2020unicon, Won2020-lb, yuan2020residual, Chentanez2018PhysicsbasedMC, Peng2021AMPAM, Fussell2021SuperTrack} has attracted increasing attention from the animation and graphics community. It has also been successfully applied in other fields, such as computer vision \cite{Yuan2021SimPoESC, Luo2021DynamicsRegulatedKP}. Originally developed by learning a single motion clip \cite{Peng2018DeepMimicED}, deep reinforcement learning (DRL) has shown impressive results in motion imitation that extend to controllable characters capable of imitating thousands of motion clips \cite{Chentanez2018PhysicsbasedMC, Won2020-lb, wang2020unicon, Luo2021DynamicsRegulatedKP}. More recent advances focus on improving the time of environmental interaction through model-based reinforcement learning \cite{Fussell2021SuperTrack}, as well as improving the use of motion clips through generative models \cite{Peng2021AMPAM} and motion warping \cite{Lee2021LearningAF}. Although these controllers can achieve impressive imitation results, they are often limited to a small subset of human motion defined by task and training data. To learn a more universal humanoid controller, residual force control \cite{yuan2020residual} is proposed to help balance the humanoid throughout the imitation process and has been successfully applied to learn controllers that can mimic up to tens of thousands of motion clips \cite{Yuan2021SimPoESC, Luo2021DynamicsRegulatedKP}. However, these methods have been focusing on obtaining better controllers \cite{Mourot2021ASO} for a \textbf{single} simulated character without considering body shape and weight variations.

\subsection{Character Design Optimization}
Automated character and agent design has been extensively explored in both the graphics and robotics community, since computer-aided designs have great potential to create elegant and effective characters or robots. Earlier work often employ evolutionary methods such as CMA-ES \cite{Luck2019DataefficientCO} to jointly optimize for better design and control policies. Similar to character control, recent advances in DRL based methods also achieved impressive results in optimizing not only the character's physical attributes such as bone length and weight \cite{Schaff2019JointlyLT, Ha2019ReinforcementLF}, but also its skeleton structure and morphology \cite{Luck2019DataefficientCO, Wang2019NeuralGE, Yuan2021Transform2ActLA}. However, most of the prior arts have been focusing on a low-dimensional, single task setting, where the character is tasked to achieve simple tasks such as "moving forward" or "hopping around" inside a simulation. Due to the high dimensionality of possible design parameters and morphology, such frameworks have yet to be applied to humanoids, where controlling a humanoid for motion imitation is an unsolved task itself. Although some exploration has been made in training controllers that can adapt to different body shape parameters \cite{Won2019-ce}, the controllers learned in this setting can simple tasks such as moving forward or imitating a single motion clip. In this work, we tackle the challenging task where we aim to discover humanoid designs that not only specialize in a given motion clip but also retain the ability to imitate general motion sequences.

\section{Approach}

 \begin{figure}[t]
    \centering
    \includegraphics[width=\linewidth]{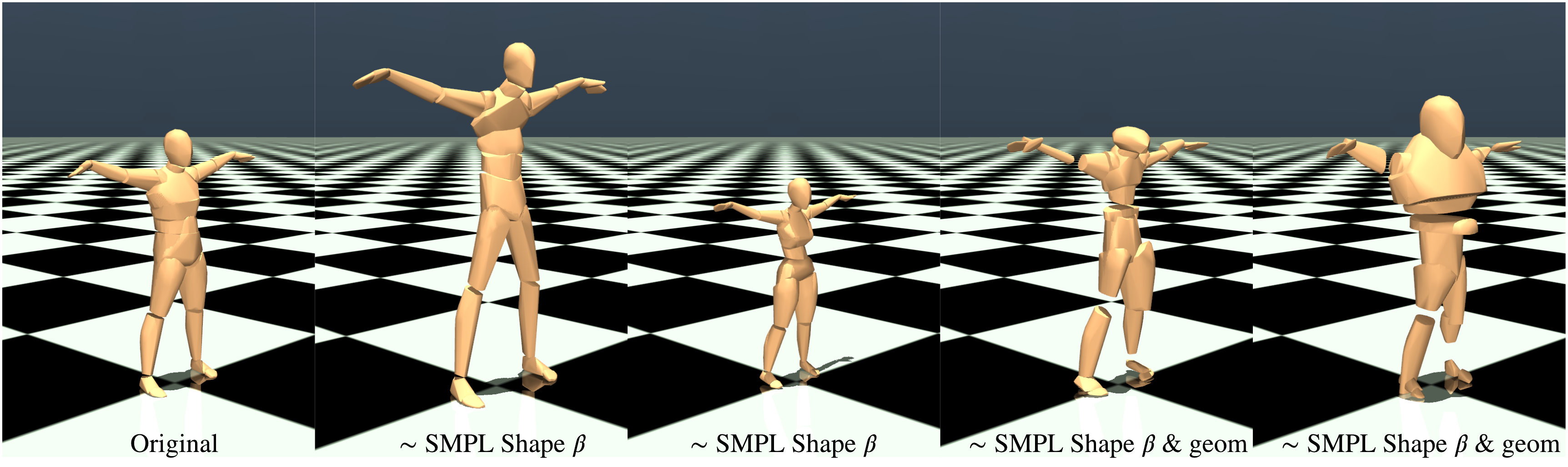}
    \caption{Visualization of the character design space our algorithm can search through. From left to right: 1) the original humanoid shape, 2)\& 3) by sampling the SMPL shape space $\beta$, 4) \& 5) by sampling the SMPL shape space and geom sizes. }
    \label{fig:design}
\end{figure}

The problem of discovering motion-dependent design can be formulated as follows: given a sequence of human motion $\bs{\widehat{q}_{1:T}}$, we aim to determine a suitable humanoid design $\bs D$ that can better imitate the given sequence. We first train a universal body shape-dependent humanoid controller (Sect.\ref{sec.uhc}) which will be fixed during the design optimization process. Then, we optimize the agent's design parameters $\bs D$ using a design-and-control framework (Sec. \ref{sec.design}). As a notation convention, we use $\widehat{\cdot}$ to denote the target pose, and normal symbols without accents to denote poses from the physics simulator (Mujoco \cite{Todorov2012MuJoCoAP}).

\subsection{Automatic Character Creation}

We first utilize an automatic simulated character creation process to generate human bodies with realistic joint limits, body proportions, and other physical properties based on the popular SMPL \cite{Loper2015SMPLAS} human mesh model. The SMPL model is a parametric human body model learned from thousands of body scans and provides a skinned mesh model of $J$ joints, $B$ bones, $V$ vertices, and a skinning weight matrix $\bs W \in \mathbb{R}^{V \times B}$. SMPL represents the human body by its shape $\bs \beta \in \mathbb{R}^{10}$ and pose $\bs \theta \in \mathbb{R}^{24 \times 3}$. Given $\bs \beta$ and $\bs \theta$, a transform function $S$ can be used to recover the location of all the mesh vertices $S(\bs \beta, \bs \theta)  \rightarrow \mathbb{R}^{V \times 3}$. By varying the shape parameter $\bs \beta \in \mathbb{R}^{10}$, we can efficiently explore a broad range of human body configurations that are within the biologically plausible range. To create convex geometries for each bone, we use the skinning weight matrix $\bs W$ similar to techniques from prior arts \cite{Christen2021DGraspPP, Yuan2021SimPoESC}. Given $\bs W$ where $\bs W_{ij}$ specifies the influence of the movement of $j^{\text{th}}$ bone to the position of the $i^{\text{th}}$ vertex, we assign vertices to bones based on the highest weight value. On the basis of this vertex-to-bone association, we can find the geometries for each bone by computing the convex hull of the assigned point cloud. Fig. \ref{fig:design} visualizes humanoids of different genders and body shapes.

\subsection{Motion imitation Policy}
\label{sec.uhc}
We first learn a single robust control policy that is capable of controlling simulated characters of different body proportions, height, gender, and weight to imitate tens of thousands of human motion sequences. Formally, we follow the standard formulation in physics-based character and model controlling a humanoid for motion imitation as a Markov Decision Process (MDP) defined as a tuple ${\mathcal M}=\langle \mathcal{S}, \mathcal{A}, \mathcal{T}, R, \gamma\rangle$ of states, actions, transition dynamics, reward function, and a discount factor. The state $\mathcal{S}$, reward $\mathcal{R}$, and transition dynamics $\mathcal{T}$ are computed by the physics simulator,  while the action $\mathcal{A}$ is given by the control policy $\pi^C_{\theta}({\bs a_t | \bs s_t, {\bs{\widehat{q}}}_t}, \bs D)$ based on the current simulation state $\bs s_t$, reference motion ${\bs{\widehat{q}}}_t$, and humanoid design attributes $\bs D$. We employ Proximal Policy Optimization (PPO)  \cite{Schulman2017ProximalPO} to find the optimal control policy $\pi^C_{\theta}$ to maximize the expected discounted reward $\mathbb{E}\left[\sum_{t=1}^{T} \gamma^{t-1} r_{t}\right]$. In the following paragraphs, we will describe in detail each component of our motion imitation framework and our enhancement to create a design-dependent humanoid controller. 

\noindent \textbf{State.} The simulation state $\bs s_t \triangleq (\bs q_t, {\bs{\dot{q}}}_t)$ consists of the character's 3D pose $\bs q_t$ and velocities ${\bs{\dot{q}}}_t$. The 3D pose of the simulated character ${\bs{q}}_t  \triangleq (\bs{q}^p, \bs{q}^r)$ consists of the character's 3D position $\bs{q}^p \in \mathbb{R}^{J \in 3}$ and orientation $\bs{q}^r \in \mathbb{R}^{J \times 3}$ for each one of the human joints. Velocity ${\bs{\dot{q}}}_t$ is defined as a concatenation of linear and angular velocities ${\bs{\dot{q}}}_t  \triangleq (\bs{\dot{q}}^p, \bs{\dot{q}}^r)$. These values encompass the full simulation state of the humanoid and are computed by the simulator at each time step.

\noindent \textbf{Policy.} The policy computes the mean of a Gaussian distribution $\bs \mu_\theta$ with a fixed covariance matrix $\bs \Sigma$: $\pi^C_{\theta}({\bs a_t | \bs s_t, {\bs{\widehat{q}}}_t}, \bs D) =\mathcal{N}(\bs \mu_\theta, \bs \Sigma)$. During training, we sample the action from the distribution to explore different actions, and during testing, we use the mean action. Given the simulation state $\bs s_t$ and reference motion ${\bs{\widehat{q}}}_t$, the policy will first use a feature extraction function $T_{E}$ to extract the difference between the simulated character state and the reference motion and transform these features into an character-centric coordinate system for computation. Specifically:

\begin{equation}
    (T_{\text{CC}}(\bs{\widehat{q}}^p, \bs{q}^p - \bs{\widehat{q}}^p, \bs{\dot{q}}_t),  (\bs{q}^r \ominus \bs{\widehat{q}}^r), \bs{\widehat{q}}^r) = T_{E}(\bs s_t, {\bs{\widehat{q}}}_t),
\end{equation}
 $T_{\text{CC}}$ transfers the input quantities from the global space into the character's local coordinate frame based on the character root joint's translation and position. Character design properties $\bs D$ will be specified in Sec. \ref{sec.design}.

\begin{figure}[t]
    \centering
    \includegraphics[width=0.8\linewidth]{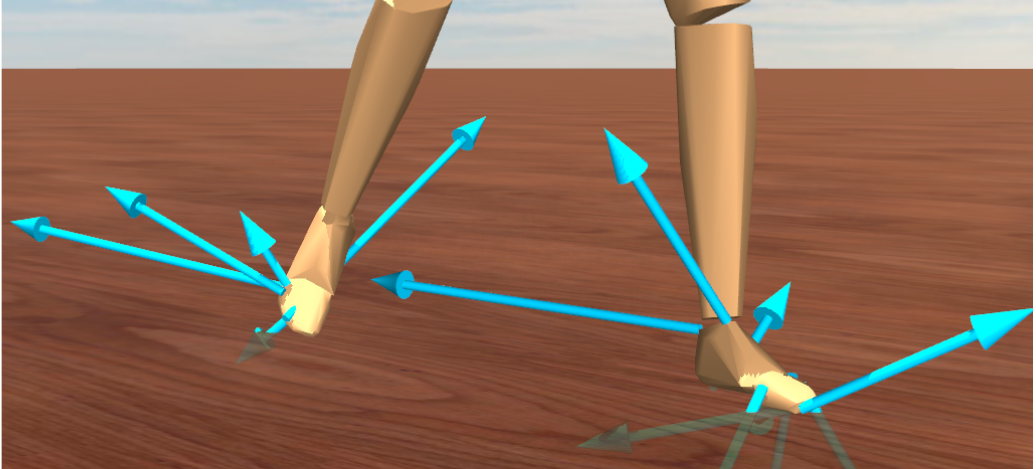}
    \caption{Residual force applied at the humanoid's foot to compensate for the dynamics-mismatch between real foot/shoes and simulation. }
    \label{fig:rfc}
    \vspace{-4mm}
\end{figure}

\noindent \textbf{Action.} The action $\bs a_t \triangleq (\bs p^d_t, \bs k_t, \bs e_t)$ consists of the target joint angles $\bs p_t$ for the PD controllers attached to each joint, PD controller gains $\bs k_t$, and the external residual force $\bs e_t$ applied to the humanoid's foot. Here we adopt the meta-PD controller \cite{Yuan2021SimPoESC} and the residual action representation \cite{Park2019LearningPP} popular in humanoid control tasks. For each joint $i$, the torque applied can be computed as ${\bs \tau}^i = {\bs k}_t^{p} \circ (\bs{p}^d_t-\bs{p}_t)- \bs{k}_t^d \circ \dot{\bs{p}}_{t}$ where $\bs{k}_t^p$ and $\bs{k}_t^d$ are learned gains by the meta-PD controller. $\bs{p}^d_t$ is the  target joint angle, $\bs{p}_t$ the current joint angle, and $\bs{\dot{p}}_t$ the current joint velocity. $\circ$ is the element-wise multiplication. Based on the observation that foot compliance and flexibility are crucial in helping the humanoid stay stable \cite{Park2018MultiSegmentFM}, we resort to applying residual force on the foot geometry to best match the ground reacting force induced by foot muscle and shoes. For each foot geometry (ankle and toe as defined in SMPL), we apply five external forces $\bs e_t \triangleq (\bs e^A_{1t}, \ldots, \bs e^A_{5t}; \bs e^T_{1t}, \ldots, \bs e^T_{5t})$ where $\bs e^A_{1t}$/$\bs e^T_{1t}$ is the first external force applied to the ankle/toe at timestep t. For each external force $\bs e^A_{1t}$, it consist of the contact point position on the geom's local coordinate frame $\bs p^A_{1t} \in \mathbb{R}^3$, its direction $\bs d^A_{1t} \in \mathbb{R}^3$, and magnitude $s^A_{1t} \in \mathbb{R}^1$. To ensure that the policy does not abuse the external force, we constrain that the external force can only be applied when the foot is in contact with other geoms (\eg the ground), as illustrated in Fig.\ref{fig:rfc}:
 
\begin{equation}
 \bs e^A_{1t}  =  
\begin{cases}
   ( \bs p^A_{1t}, \bs d^A_{1t},  s^A_{1t}),& \text{if geometry A is in contact with the ground }\\
    0 ,& \text{otherwise}
\end{cases}
\end{equation}

\noindent \textbf{Reward.}
We use the standard motion tracking reward based on comparing the simulated character state $\bs s_t$ and the input reference motion $\bs{\widehat{q}_t}$:

\begin{equation}
\begin{aligned}
    r_t &= w_p * r_p + w_v * r_v + w_e * r_e  + w_{\text{vf}} * r_{\text{vf}}\\ 
    r_p &= \text{exp}[-2.0\| \bs q^r_t \ominus {\bs{\widehat{q}_t^r}} \|^2], \text{character rotational pose reward}\\
    r_v &= \text{exp}[-0.005\| \bs{\dot q_t} - \bs{\widehat{\dot q_t}} \|^2],  \text{humanoid velocity reward}\\
    r_e &= \text{exp}[-5\| \bs q^p_t \ominus {\bs{\widehat{q}_t^p}} \|^2], \text{character positional pose reward}\\
    r_{\text{vf}} &= \text{exp}[- \| \bs e_t \|^2], \text{residual force reward}\\
\end{aligned}
\end{equation}
where $w_p, w_v, w_e, w_{\text{vf}}$ are the respective reward weights.

\noindent \textbf{Training procedure.}
We train on the training split of the AMASS dataset \cite{Mahmood2019AMASSAO} and remove motion sequences that involve human-object interactions (such as stepping on stairs or leaning on tables). This results in 11402 high-quality motion sequences. At the beginning of each episode, we sample a fixed-length sequence (maximum length 300 frames) to train our motion controller. We use reference state initialization \cite{Peng2018DeepMimicED} to randomly choose the starting point for our sampled sequence and terminate the simulation if the simulated character deviates too far from the reference motion $\bs{ \widehat{q}^p_t}$. To choose which motion sequence to learn from, we employ a hard-negative mining technique based on the success rate of the sequence during training. For each motion sequence $\bs{\widehat{Q^i}} = {\bs{\widehat{q}_0}, \ldots, \bs{\widehat{q}_T}}$, we maintain a history queue (of maximum length 50) for the most recent times the sequence is sampled: ${h^i_1, \ldots h^i_{50}}$, where each $h^i_0$ is a Boolean variable indicating whether the controller has successfully imitated the sequence during training. We calculate the expected success rate $s^i$ of the sequence based on the exponentially weighted moving average using the history $s^i = \text{ewma}(h^i_1, \ldots h^i_{50})$. The probability of sampling sequence $i$ is then $P\left(\widehat{\boldsymbol{Q}}^{i}\right)=\frac{\exp \left(-s^{i} / \tau\right)}{\sum_{i}^{J} \exp \left(-s^{i} / \tau\right)}$ where $\tau$ is a temperature parameter. Intuitively, the more we fail at imitating a sequence, the more likely we will sample it. Each $s^i$ is initialized as 0.

\begin{figure*}[t]
    \centering
    \includegraphics[width=\linewidth]{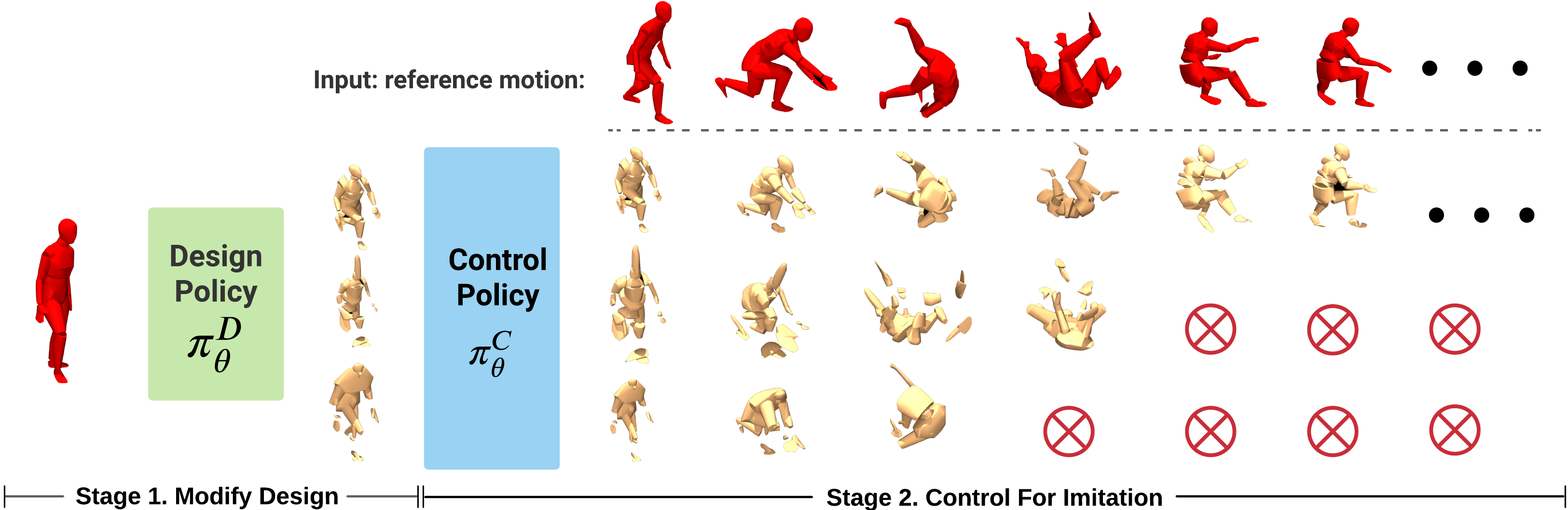}
    \caption{Given a reference motion sequence, our framework first uses a design policy $\pi^D_{\theta}$ to sample potential candidates, and then uses a control policy $\pi^C_{\theta}$ to imitate reference motion to gather feedback (reward) about the designs.}
    \label{fig:pipeline}
\end{figure*}

\subsection{Motion-dependent Character Discovery}
\label{sec.design}
To discover suitable character designs based on a sequence(s) of similar motions, we employ a two-stage design and control framework. While prior arts \cite{Yuan2021Transform2ActLA, Wang2019NeuralGE} apply similar techniques in finding simple characters for the task of "moving forward", we focus on discovering designs that are capable of specializing in an input sequence while retaining its ability to perform diverse activities. We achieve this by pairing a humanoid controller with an additional design policy and optimizing the agent design through first-order optimization. Instead of training both the control and design policy jointly from scratch, which can be unstable and time-consuming, we utilize a pretrained humanoid controller and freeze its weights during the design optimization process. This way, the pretrained controller acts as a human dynamics prior that constrains the design policy to only find agents that are controllable yet more suitable for imitating the specified motions. 

\noindent \textbf{Agent Design Parameters.} The design parameters are defined as: $$\bs D \triangleq (\bs \beta, w, h, \bs b, \bs f, \bs m, \bs g).$$ Here we modify the humanoid's SMPL shape parameter $\bs \beta$, weight $w$, height $h$, joint frictionloss $\bs f$ (joint dry friction)  \cite{Todorov2012MuJoCoAP} , the size and density of each bone's geometry $\bs m$, and motor gear for each joints' actuator $\bs g$. These parameters control the simulation properties of the character such as mass, inertia, and motor strength. Fig. \ref{fig:design} shows possible humanoid designs by sampling the SMPL shape parameters and sizes for each bone's geometry.

\noindent \textbf{Dual Policy for design and control.}
Given a pretrained control policy $\pi^C_{\theta}({\bs a_t | \bs s_t, {\bs{\widehat{q}}}_t}, \bs D)$, we augment it with a design policy $\pi^D_{\theta}$ that aims to maximize the accumulated reward conditioned on the pretrained controller: 
\begin{equation}
    \bs D^* = \underset{D}{\arg \max } \quad \mathbb{E_{\bs D, \pi^C_\theta}}\left[\sum_{t=1}^{T} \gamma^{t-1} r_{t}\right],
\end{equation} 
Design policy $\pi^D_{\theta}({\bs D | \bs s_0, {\bs{\widehat{q}}}_0})$ is conditioned on the pose of the first frame of the input reference motion ${\bs{\widehat{q}}}_0$ and computes the optimal design for the input sequence. Each episode is divided into two stages: design and control. At the beginning of each episode, we first use the design policy to change the agent's physical characteristics $\bs D$, and then roll out the control policy for motion imitation. Combining the two policies, we form the overall decision-making framework $\bs {\pi_\theta}$:

\begin{equation}
 \bs {\pi_\theta} ({\bs{a^d_t}, \bs D | \bs s_t, {\bs{\widehat{q}}}_t}, \bs D_t)  =  
\begin{cases}
    \pi^D_{\theta}({D | \bs s_0, {\bs{\widehat{q}}}_0}), \text{ design stage}\\
    \pi^C_{\theta}({\bs a_t | \bs s_t, {\bs{\widehat{q}}}_t}, \bs D), \text{ control stage},
\end{cases}
\end{equation}

The design and control algorithm can be seen in Fig. \ref{fig:pipeline} and Algorithm~\ref{alg:design_opt}. We use the same reward and optimization algorithm (PPO) as in the motion imitation case, since the main objective remains the same: better motion tracking. Notice that the agent does not receive any reward ($r_0 = 0$) at the design stage since interacting with the environment while changing the character's design is undesirable. Rather, the reward signal for the design policy comes from the control policy's interaction with the environment. We learn a new value function conditioned on the agent design to find the best design parameters:
\begin{equation}
    \mathcal{V}(\bs s_t, \bs D) \triangleq \mathbb{E}_{D, \pi_{\theta}^{C}}\left[\sum_{t=1}^{T} \gamma^{t-1} r_{t}\right].
\end{equation}

\begin{algorithm}[t]
\small
\caption{Character Design Discovery via Design-and-Control Optimization}\label{alg:design_opt}
\textbf{Input:} pretrained humanoid control policy ${\pi^C_{\theta}}$, target motion sequence(s) $ \bs{\widehat Q}$

\While{\text{not converged}}{
     $\mathcal{M} \gets \emptyset$ \tcp*[f]{initialize sampling memory}  \,\; 
    \While{$\mathcal{M}$ not full}{  
        $D_0 \gets$ initial humanoid design \,\;
        $\bs{\widehat{q}_{1:T}} \gets$  random sequence of motion from dataset $\bs{\widehat Q}$ \,\;
        \For(\tcp*[f]{Design Stage}){$t = 1$}{
            
            $\bs D \sim \pi^D_{\theta}({D | \bs s_0, {\bs{\widehat{q}}}_0})$
            sample design action \,\;
            $r_t \gets 0$; store $(r_t, \bs D)$ into $\mathcal{M}$\,\;
        }
        $\bs s_0 \gets$ initialize simulation state based on design $\bs D$\,\; 
        \For(\tcp*[f]{Control Stage}){$t = 2, \ldots, T$}{
            $\bs a_t \sim \pi^C_{\theta}({\bs a_t | \bs s_t, {\bs{\widehat{q}}}_t}, \bs D)$ sample control action from pretrained controller\,\;
            $ \bs s_{t+1} \gets \mathcal{T}(\bs s_t | \bs s_t, \bs a_t)$  simulation dynamics \,\;
            $r_t \gets$ imitation reward; \,\;
            store $(r_t, a_t, \bs s_{t+1}, {\bs{\widehat{q}}}_t,  \bs D)$ into $\mathcal{M}$\,\;
        }
  }
  update $\pi^D_\theta$ with PPO using collected samples from $\mathcal{M}$ 
}
\Return{$\pi^D_\theta$ and $\bs D$}
\end{algorithm}

\section{Experiments}
We conduct our experiments on the motion imitation task and try to gauge the effectiveness of 1) the general humanoid controller for the general motion imitation task and 2) our humanoid design discovery process's ability to find suitable character designs based on input motion sequence(s) of different categories and requirements. To test our motion imitation framework, we report motion imitation results on both the training and testing split of the AMASS dataset. For agent design discovery, we report results on finding characters for 1), finding the best design for single sequences (parkour, karate, belly dancing, boxing), 2) for a category of sequences (dancing and running) and 3) for the whole AMASS training split. 

\subsection{Metrics}
We report motion tracking quality through a suite of pose-based and physics-based metrics:

\begin{itemize}
\item Mean per joint position error: $ \bf E_{\text{mpjpe}}$ (mm) is a popular 3D human pose estimation metric and is defined as $\frac{1}{J}\| {\bs q}^{\text{p}} - \widehat{\bs q}^{\text{p}} \|_2$ for $J$ number of joints. This value is root-relative and is computed after setting the root translation to zero.

\item Mean per joint position error-global: $ \bf E_{\text{mpjpe-g}}$ (mm) this value computes the joint position error in global space without zeroing out the root translation. It better reflects the overall quality of motion tracking. 

\item Acceleration error: $\text{E}_{\text{acc}}$ (mm/frame$^2$) measures the difference between the ground truth and estimated joint position acceleration: $\frac{1}{J}\| \ddot{\bs q}^{\text{p}} - \widehat{\ddot{\bs q}}^{\text{pos}} \|_2$.

 \item Success rate: $\text{S}_{\text{succ}}$ measures whether the humanoid has fallen or deviates too far from the reference motion during the sequence. If the humanoid falls during imitation, we will reset the simulation based on the reference motion at the timestep of failure. 
\end{itemize}
\begin{table*}[h]
\caption{Evaluation of agent specializing in various challenging tasks. Here the suffix indicates the number of motion sequences fed into the alogrithm.}
\raggedleft
\resizebox{0.9\linewidth}{!}{%
    \begin{tabular}[b]{l|rrrr} 
    \midrule
    Sequence  & $\text{S}_{\text{succ}} \uparrow$ & $\text{E}_{\text{mpjpe}}\downarrow$ & $\text{E}_{\text{mpjpe-g}} \downarrow$   & $\text{E}_{\text{acc}} \downarrow$     \\ \midrule
    Belly Dance-1  &  ${0\%} \rightarrow \textbf{100\%}$ &  ${183.3} \rightarrow \textbf{36.9}$ &  ${347.1} \rightarrow \textbf{55.0}$ &  ${7.6} \rightarrow {7.6}$ \\  
    Parkour-1  & ${0\%} \rightarrow \textbf{100\%}$ &  ${175.6} \rightarrow \textbf{87.1}$ &  ${324.2} \rightarrow {146.4}$ &  ${24.8} \rightarrow \textbf{15.0}$ \\  
    Karate-1  &  ${100\%} \rightarrow {100\%}$ &  ${35.9} \rightarrow \textbf{30.1}$ &  ${45.8} \rightarrow \textbf{39.9}$ &  ${7.1} \rightarrow {7.1}$ \\  
    Crawl-1  &  ${100\%} \rightarrow {100\%}$ &  ${66.4} \rightarrow \textbf{40.6}$ &  ${307.6} \rightarrow \textbf{67.2}$ &  $\textbf{3.8} \rightarrow {4.3}$ \\  
    Cartwheel-1  & ${0\%} \rightarrow \textbf{100\%}$ &  ${160.9} \rightarrow \textbf{37.4}$ &  ${284.8} \rightarrow {66.2}$ &  ${8.1} \rightarrow \textbf{4.9}$ \\  
    \hline
    Dance-200  & ${57.0\%} \rightarrow \textbf{72.0\%}$ &  ${84.1} \rightarrow \textbf{58.0}$ &  ${146.7} \rightarrow \textbf{98.7}$ &  ${13.3} \rightarrow {13.3}$ \\  
    Tennis-60 &  ${96.7\%} \rightarrow \textbf{100\%}$ &  $\textbf{27.8} \rightarrow {21.8}$ &  ${40.5} \rightarrow \textbf{30.7}$ &  ${4.2} \rightarrow \textbf{4.1}$ \\
    Crawl-37  &  ${94.6\%} \rightarrow \textbf{94.6\%}$ &  ${62.0} \rightarrow \textbf{40.6}$ &  ${163.6} \rightarrow \textbf{87.9}$ &  ${5.6} \rightarrow {9.1}$ \\
    Cartwheel-4  &  ${25\%} \rightarrow \textbf{75\%}$ &  ${219.6} \rightarrow \textbf{89.4}$ &  ${393.6} \rightarrow \textbf{166.1}$ &  ${19.4} \rightarrow \textbf{12.5}$ \\
    Kick-302 &  ${98.3\%} \rightarrow {98.3\%}$ &  ${45.5} \rightarrow \textbf{38.8}$ &  ${75.4} \rightarrow {62.4}$ &  $\textbf{7.5} \rightarrow {9.1}$ \\
    \bottomrule 
    \end{tabular}
    }\\ 
\label{tab:design}
\end{table*}

\begin{table}[h]
\caption{Evaluation of motion imitation for our humanoid controller using target motion from the train and test split of the AMASS dataset.}
\raggedleft
\resizebox{1\linewidth}{!}{%
    \begin{tabular}[b]{l|rrrr|rrrr} 
    \toprule
    \multicolumn{1}{c}{} & \multicolumn{4}{c|}{AMASS dataset Training (11402 sequences)}  &  \multicolumn{4}{c}{AMASS Testing (140 sequences) } 
    \\ 
    \midrule
    Method  & $\text{S}_{\text{succ}} \uparrow$ & $\text{E}_{\text{mpjpe}}\downarrow$ & $\text{E}_{\text{mpjpe-g}} \downarrow$   & $\text{E}_{\text{acc}} \downarrow$ & $\text{S}_{\text{succ}} \uparrow$ & $\text{E}_{\text{mpjpe}}\downarrow$ & $\text{E}_{\text{mpjpe-g}} \downarrow$   & $\text{E}_{\text{acc}} \downarrow$    \\ \midrule
    No-RFC  &  {89.7\%} &  {32.1} &  {50.7} &  {11.0} &  {65.5\%} &  {86.6} &  {156.1} &  {25.4}\\  
    RFC-Root &   {94.7\%} &  {34.5} &  {50.9} & \textbf{4.5} & {80.7\%} &  {51.1} &  {76.2} &  {12.5} \\
    RFC-Foot &  \textbf{95.6\%} &  \textbf{23.8} &  \textbf{36.5} &  {4.71} & \textbf{91.4\%} &  \textbf{35.3} &  \textbf{60.1} &  \textbf{10.5} \\
    \hline
    RFC-All  &  {99.9\%} &  {17.8} &  {24.3} &  {4.1} & {99.3\%} &  {24.4} &  {36.4} &  {9.1}\\  
    \bottomrule 
    \end{tabular}
    }\\ 
\label{tab:uhc_amass}
\end{table}

\subsection{Motion Imitation}
Table \ref{tab:uhc_amass} shows the quantitative results of our controller on motion imitation. Here we report the results of UHC with different configurations: UHC w/o RFC is our controller without using any external residual force; UHC-RFC-Root is similar to \cite{Luo2021DynamicsRegulatedKP} where the external force is applied at the root of the character; UHC-RFC-Foot is our proposed method where the external force is applied on the foot and only when the foot is in contact with the ground; UHC-RFC-All is a configuration where we apply an external force to every joint in the human body. UHC-RFC-All can serve as an oracle method showcasing the best performance of a physics-based humanoid controller, where imitation quality rather than full physical realism of the simulated character is the priority. As can be shown in the result, our proposed method outperforms the previous state-of-the-art in terms of motion imitation quality on all metrics and is close to the physically implausible UHC-RFC-All case. Compared to UHC-No-RFC, we can see that our external force on the foot is effective compensating for the foot instability induced by the simple foot design (as indicated by the high acceleration error, the humanoid is jittering to try to stay upright). When UHC-RFC-Root and UHC-RFC-Foot are compared, it is clear that applying residual force only on the foot is more effective and has a more intuitive justification than using a supporting force on the character's root.

\subsection{Character Design Discovery}
We evaluate our character design discovery framework by assigning it to find suitable humanoids for challenging motion sequences (s) such as belly dancing, parkour, karate, cartwheeling, etc. Such sequences are highly dynamic and often require the human performer to have a special physique to properly conduct these moves. For example, an expert dancer will be well balanced, while a crawler may have more prominent elbows. As a result, our vanilla humanoid controller may fail at imitating such sequences, and modifying the agent's design can improve our controller's performance. Our evaluation is conducted using two different modes: single sequence and multiple sequence discovery, where we find designs that best fit one sequence or a category of motion sequences.

\begin{figure*}[t]
    \centering
    \includegraphics[width=\linewidth]{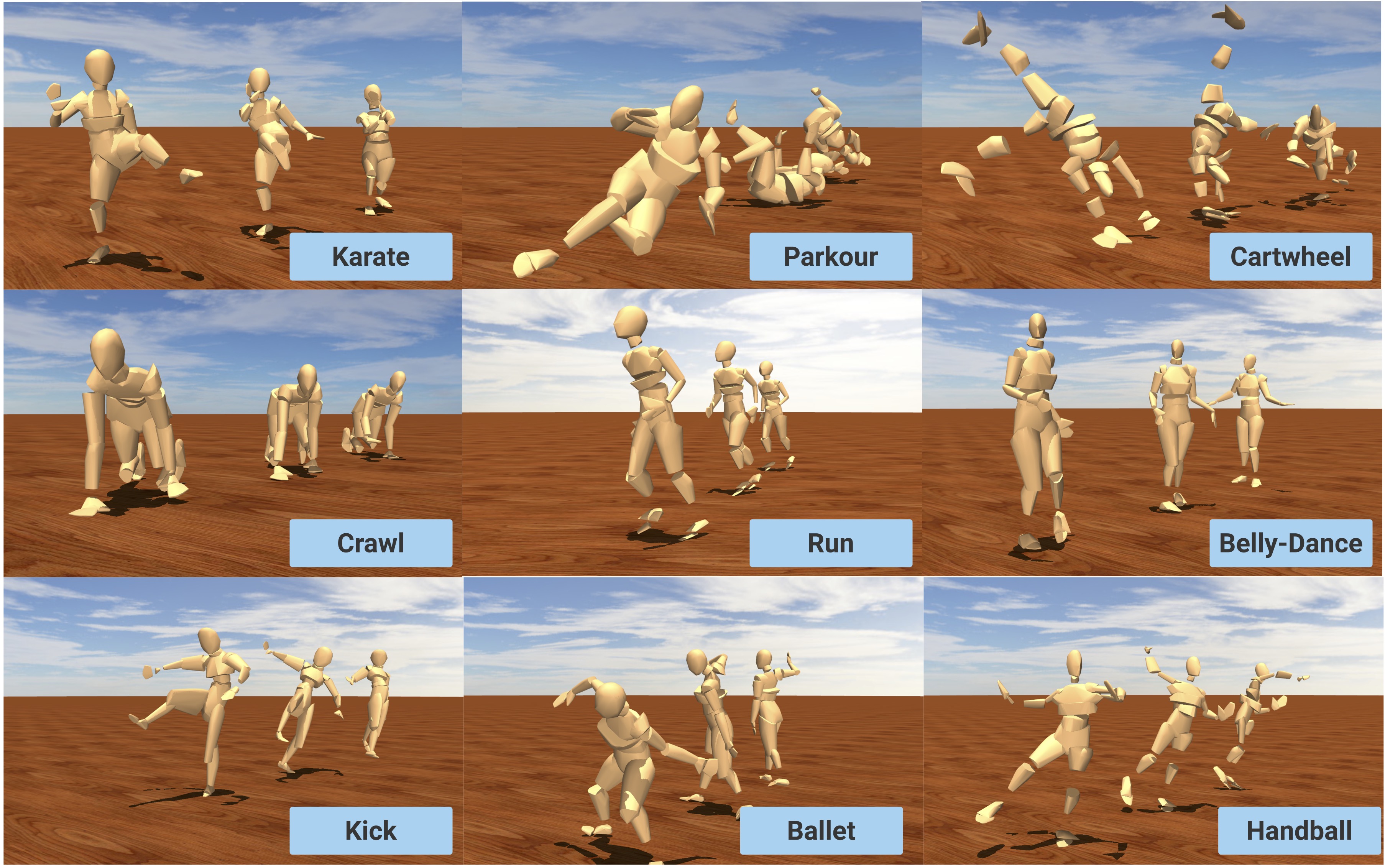}
    \caption{Humanoid designs for various kinds of challenging motions.}
    \label{fig:main}
\end{figure*}

\noindent \textbf{Single Sequence.}
The top half of Table \ref{alg:design_opt} shows our results in finding the character design that fits the \textit{best single sequence}. From the metrics, we can see that for each individual sequence, optimizing the design parameters can significantly improve the imitation quality, and sometimes give the controller the ability to successfully imitate a sequence without falling. Notice that our controller is frozen after training on the AMASS dataset and that only the design parameters of the simulated character are altered. This shows that the parameters we search for are essential in the motion tracking process, and tuning them can result in better motion imitation quality. Upon visual inspection in Fig \ref{fig:main}, we can see that our recovered agents are aesthetically pleasing and fit our intuitive understanding of each motion category: the parkour agent acquires wider hips and thighs for better ground contact, the cartwheeling agent obtains bigger hands for stable support, while the karate player becomes more stout for better balance. 

\noindent \textbf{Category-level.}
On the category level in Table \ref{tab:design}, we can observe similar trends as the single sequence case. Our design optimization process is able to find suitable agents for a whole category of motion sequences, demonstrating its ability to generalize to a suite of diverse motions that shares similar traits. As motion is best seen in videos, we refer our reader to our \href{ supplementary video}{https://youtu.be/uC0P2iB56kM} for visual results. 

\noindent \textbf{Test set transfer.}
One of our main goals for fixing the control policy is to discover agents that can perform specific actions in certain motion categories while preserving the agent's ability to imitate general motions (such as walking and running). To this end, we evaluate these special agent's performance on the AMASS test set. Results from table \ref{tab:transfer} shows that the specialized agents, while becoming better at imitating one category of motions, still retain their ability to perform general motion and largely preserve their performance on the AMASS test split.

\begin{table}[h]
\caption{Performance of our special humanoid designs on the AMASS test split}
\resizebox{0.8\linewidth}{!}{%
    \begin{tabular}[b]{l|rrrr} 
    \toprule
    \multicolumn{1}{c}{} & \multicolumn{4}{c}{AMASS Test Set}
    \\
    \midrule
    Dance-200  &  \textbf{92.8\%} &  {40.4} &  {70.4} &  {12.1}\\  
    Tennis-60  &  {89.9\%} &  {33.9} &  {55.7} &  {11.9}\\  
    Crawl-37  &  {86.3\%} &  {36.7} &  {54.4} &  {13.0}\\  
    Cartwheel-4 &  {89.9\%} &  {36.8} &  {64.9} &  {11.8}\\  
    Kick-203  &  {90.6\%} &  {52.9} &  {92.9} &  {14.6}\\  
    \midrule
    RFC-Root & {91.4\%} &  \textbf{35.3} &  \textbf{60.1} &  \textbf{10.5} \\
    \bottomrule 
    \end{tabular}
    }\\ 
\label{tab:transfer}
\end{table}

\section{Limitations and Future Work}
Currently, our agent discovery process is still relatively time-consuming: utilizing a pretrained controller, it takes around 3 hours to find suitable agents for a single motion sequence (longer for more diverse motions). Our network is also unimodel and can only recover a single agent design regardless of the number of input motion sequences. In the future, we aim to expand our agent design policy to incorporate richer sequence-level information and enable our framework to generate diverse designs based on input motion at test time without additional training. 

\section{Conclusion}

In this work, we present an automatic character design framework for generating controllable agents for various categories of highly dynamic human motion. Given a target motion sequence (either through MoCap,  keyframe animation, or recovered from video \cite{Peng2018SFVRL}), our pipeline can recover humanoid designs that can specialize on such sequences while retaining the ability to imitate diverse motion sequences. Notably, we also propose a more phsycially valid humanoid controller than prior art, and compensate for the lack of biologically sound foot design through residual force control. Recovered agents can serve as inspiration for creating physically valid controllable characters based on unique motion requirements.

\bibliographystyle{ACM-Reference-Format}
\bibliography{sample-bibliography}

\appendix

\end{document}